# A Design and Development of Rubrics System for Android Applications

Kaustubh Kundu[*],  Sushant Yadav, Tayyabbali Sayyad
*Dept of Information Technology, Don Bosco Institute of Technology , Kurla(W) , Mumbai 400070, India*
*Corresponding author: Kaustubh Kundu*

**ABSTRACT: -** Online grading systems have become extremely prevalent as majority of academic materials are in the process of being digitized, if not already done. In this paper, we present the concept of design and implementation of a mobile application for 'Student Evaluation System', envisaged with the purpose of making the task of evaluation of students performance by faculty and graders facile. This application aims to provide an user-friendly interface for viewing the students performance and has several functions which extends the Rubrics with graphical analysis of students assignments. Rubrics evaluation system is the widespread practice in both the software industry and the educational institutes. Our application promises to make the grading system easier and to enhance the effectiveness in terms of time and resources. This application also allows the user/grader to keep track of submissions and the evaluated data in a form that can be easily accessed and statistically analysed in a consistent manner.
---
Date of Submission: 15-03-2018                                                                     Date of acceptance: 30-03-2018
---

## I. INTRODUCTION

In the past decades, mobile phones were equipped with very basic capabilities. These were devices which were only capable of making calls and sending text messages through wireless connections. Due to the recent advancement in technologies, capabilities of mobile phones have reached beyond expectations. In today's society, cellular phones have made a significant impact on people's lives, owing to the advancements in computing speed, power and associated hardware. Different operating systems have also been developed like Android, iOS, Windows and Symbian. These operating systems have allowed developers to make mobile applications using various programing platforms. Utilizing these platforms, developers have taken the opportunity to create and develop applications. It is believed that owing to these advancements in mobile hardware, as well as the increase in the penetration of mobile telephony services into hitherto untapped portions of the society, consumers have started availing services which was previously unavailable to them. As we move towards a digitized society, it becomes imperative to develop applications which will help in taking advantage of these services. There is also a prediction that education based applications will also increase due to the access of smartphones especially in rural areas [1]. Developing applications on education, therefore, have become at most importance for the developers to greater ease of student-teacher interaction.

Here we present the development of a Rubrics system for the purpose of students evaluation. The system design and the methodology adapted in this development are described. To make the evaluation system interactive, various facilities to view and manage the storage data included in this system are elaborated. Statistical analysis of the incorporated data to get the overview of the students performance is also presented.

## II. FAMILIARISATION OF THE RUBRICS SYSTEM

The chronological development of the mobile technologies in the past decade is depicted in Fig.1 [2]. Initially the mobile phones were 1st generation (1G) wireless technology, which was analogue to cellular phones those transmitted and received only voice signals using NMT, AMPS mobile communications. 2G, known as the second generation wireless technology, was the first digital fidelity cellular phones, including voice and data signals with GSM, CDMA, TDMA communication protocol. 3G was the third generation wireless technology, which combined video telephony/internet surfing on cellular phones with voice, data and video signals using W-CDMA, UMTS communication protocol. Currently 4G, the fourth generation mobile technology is high speed IP based cellular phones with voice, internet and data signals.

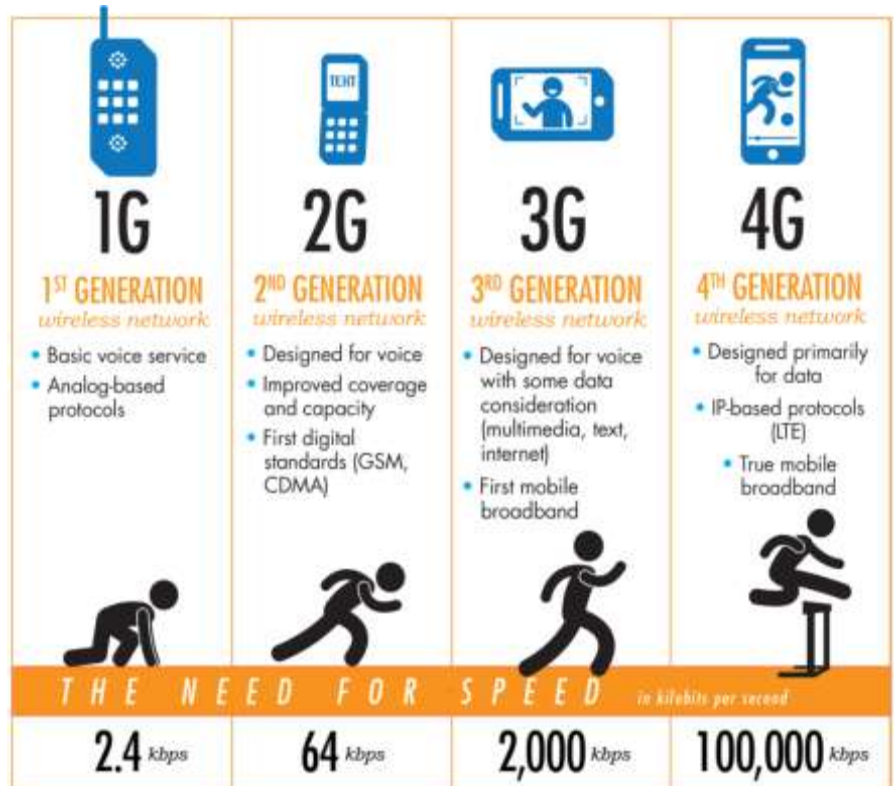

**Fig. 1. The history of mobile devices.**

Rubrics is an important tool that helps evaluators to analyse the performance of participants. In order to understand about the Rubrics system, we first need to understand why, when and where it is used. As shown in Fig. 2, the Rubrics system can be used in several spheres, including school, colleges and government offices in order to evaluate and analyse the performance of the students and employees. These factors help the evaluator to evaluate and analyse the performance in terms of the respective strong and weak points and provide transparency to those getting evaluated based on the criteria that is set up by the evaluator. Although, the Rubrics has wide applications in various fields, this is the platform which is hitherto unexplored in android. Till date, the similar developed systems available in the literature [3,4,5,6] are lacking of user friendliness, creation of Rubrics, unavailability of statistical analysis and student feedback. Hence, these anomalies need to be addressed for the development of the students evaluation system.

In our work, this mobile-based platform for the graders, which will evaluate the work according to the designed metric developed by the grader. This system displays all the performance of the students in a graphical format so that the grader will get a clear idea about where the studentsare lagging and can improve upon. Using this Rubrics system, the grader can review the student's work as well as give proper feedback to the student and ensuring that the student gets one-to-one attention from the grader.

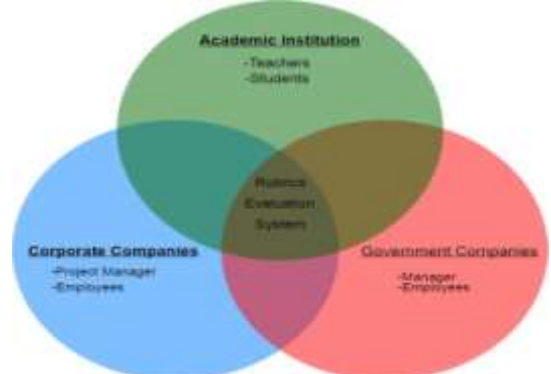

**Fig. 2.Places where Rubrics can be used**

## III. PROPOSED SYSTEM DESIGN AND METHODOLOGY
*A. System Architecture*

The following architectural diagram given in Fig. 3 shows the different modules that make up the Rubrics system for android application. There areplenty of advantages over existing systems. These include, the obvious benefits such as minimal or no pricing, coupled with a user friendly system that can be synced in the offline mode. Further, the multiple built-in templates are dynamic and offer a graphical performance analysis along with an interactive feedback to the students.

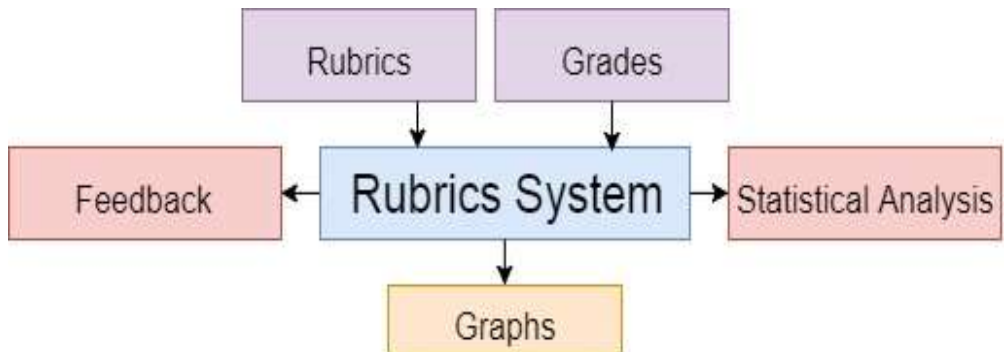

**Fig. 3. Android Architectural Framework and Module development.**

The proposed android architectural framework and module development encompasses fourmodules namely:
 (1) Rubrics (2) Grades (3) Statistical Analysis (4) Graphs and (5) Feedback.

*(1) Rubrics*
In this module the user can create his/her own Rubrics by specifying the parameters based on the type of assessment he/she wants to carry out such as assignments, exams and so on. It can also make use the predefined Rubrics.

*(2) Grades*
In this module the user can grade the students on choosing the Rubrics created by the user or the predefined rubrics.

*(3) Statistical Analysis*
This module provides a representation of the analysis done by the user. It describes the nature of the analysis. It consist of:
i) Mean -  That shows the average of the data.
ii) Median  - That shows the central tendency of the data.
iii) Mode - That shows the data values that appear the most.

*(4) Graphs*
This module presents the assessment of the user in a graphical representation of the whole class. It consists of two parts
i) Total marks - It displays the total marks of the student in a Bar graph.
ii) Threshold marks - It displays the data in a pie chart which shows how many students are below and above a certain threshold marks as set by the user.

*(5) Feedback*
This module helps the user send the marks to the students via email system in the pdf or excel format.

*B. Calculations of the Rubrics system:*

In this system the main part of calculation is done in central tendency, which is a part of statistical analysis. We have considered including central tendency because it represents a single value that attempts to describe the whole set of data which in our case alludes to the grades of all the students by identifying the central position. Therefore they are dubbed as summary statistics. Mean, median and mode are considered to be the valid measures of central tendency.

*i) Mean :* It is the most popular and well known measure in the central tendency. The mean is equal to the sum of all the values divided by the number of values in the data set. In our case it is equal to the sum of all the grades divided by the number of grades in the data set. The main advantage of implementing this is that it can be used in both continuous and discrete data.

$$\bar{x} = \frac{(x_1 + x_2 + \cdots + x_n)}{n}$$

*ii) Median* : It is the middle score of the set of data that has been arranged either in ascending or descending order of magnitude. In our case it is the middle value that will be appearing in the dataset of grades. But there can be two possibilities that can occur while calculating median.

Case I - If the dataset if of odd number of values, then the middle most value will be selected.
Case II - If the dataset is of even number of values, then the mean of the two middle numbers is chosen.

*iii) Mode* : It is the most number of times a certain value occuring in a particular dataset. In our case the most number of times it is the particular mark occurring the most number of times in a dataset of grades.

*C. System Components*
 Fig. 4 represents the overall system components considered in this Rubrics system

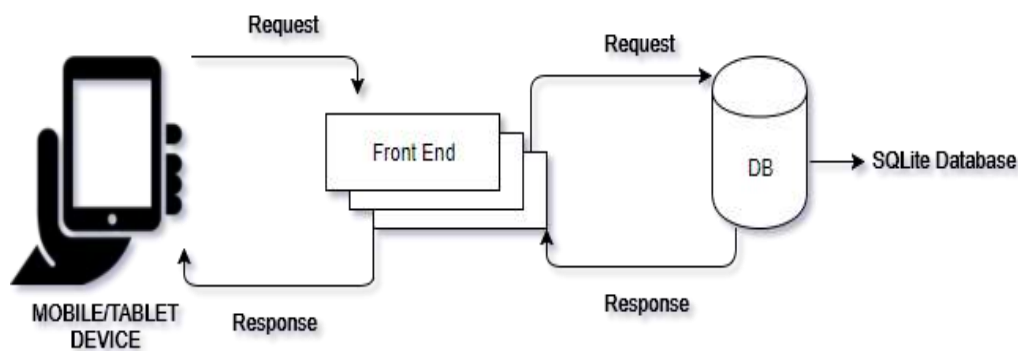

**Fig. 4. Overall System components.**

*(1) Front end*
Front end is the visual part of the application with which user interacts. In this system the grader is provided with user friendly interface through which they can provide data to the system.

*(2) Database Server*
This component hosts the database which is used for storing the data provided by the user from the front end. The database usedhere is Sqlite mobile database. It is a relational database which stores data to a text file on a device. Android comes with built in Sqlite database implementation.
As shown in Fig. 5, system wasdesigned based on the Android Operating System. This mobile application includesmanage class, manage assignment, manage Rubrics, start grading, view graphs and attendance. There is also an important component called the provider service. Its main function is data processing. However, both parts require Client server protocol for interfacing.

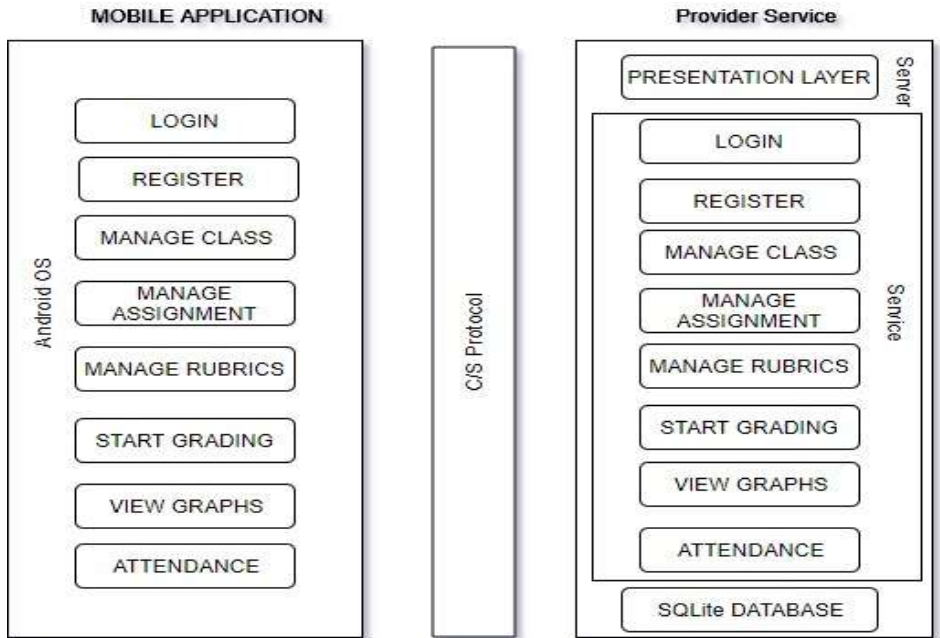

**Fig. 5. The conceptual framework of the recommended exercise system**

After simulating number of probable scenarios, the system design is conceptualised. In this design phase the system process diagram is presented in Fig. 6.

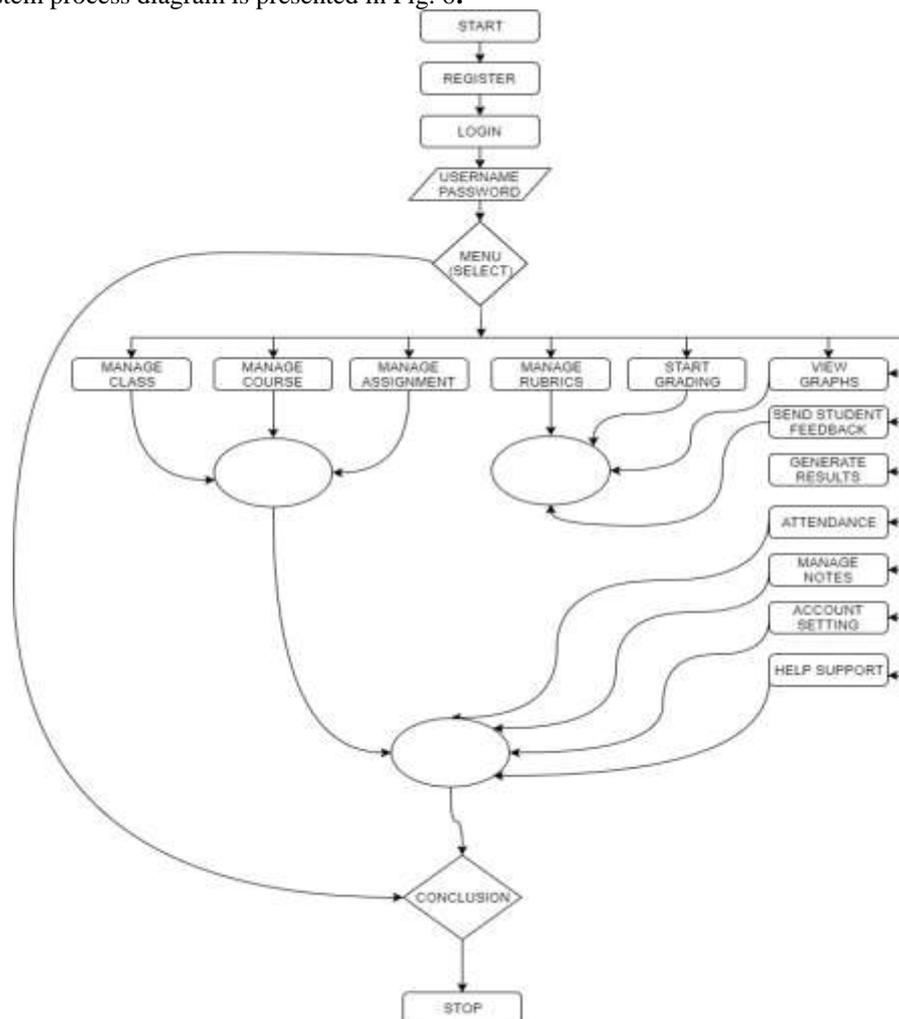

**Fig. 6. The system process diagram.**

## IV. RESULTS

Following the process flow diagram in Fig. 6, the software based on Android OS was developed. The overview of the probable exercise system for the user interface is tabulated in Fig. 7. These are mainly manage class, manage assignment, manage Rubrics, start grading, view graphs and attendance. A glimpse of the actual pages developed on the mobile screen are shown in Fig. 8. As shown, it becomes easy for the programmer to communicate with the stakeholders about each display that should be shown to users by the system.

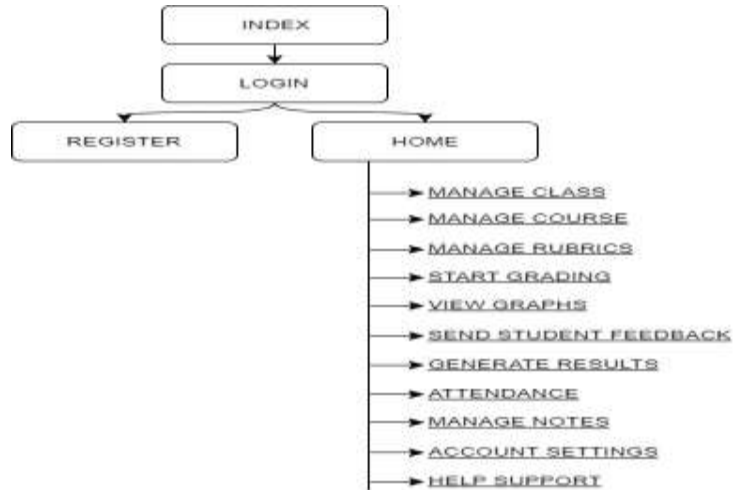

**Fig. 7. Overview of the recommended exercise system.**

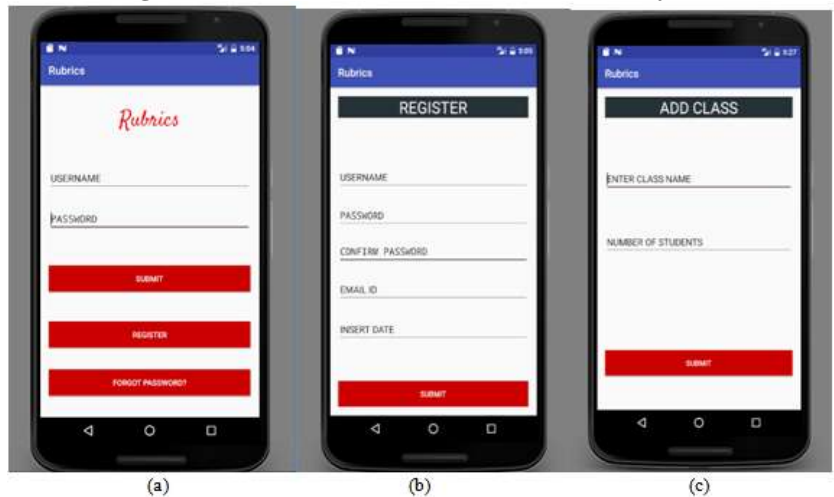

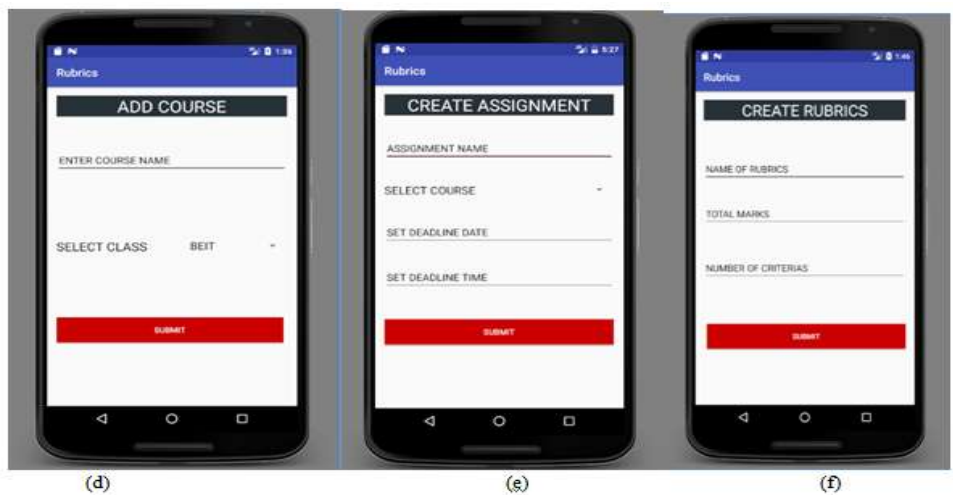

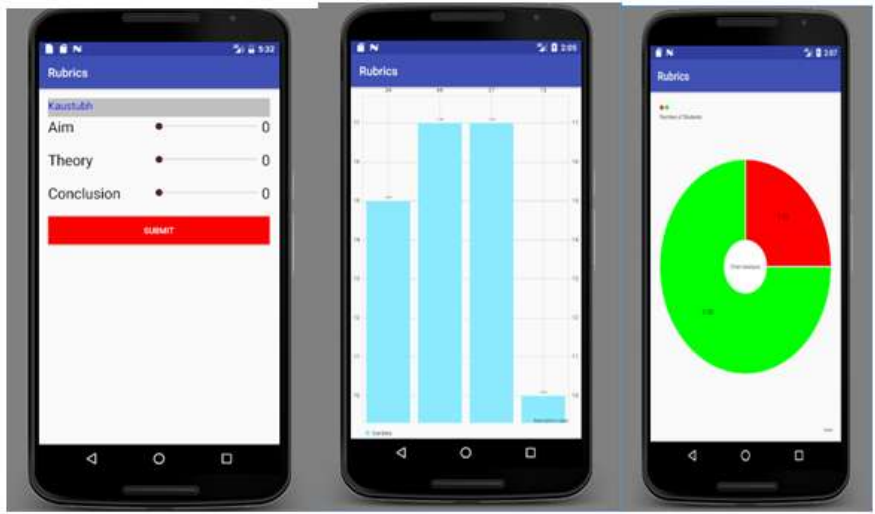

(g)       (h)       (i)

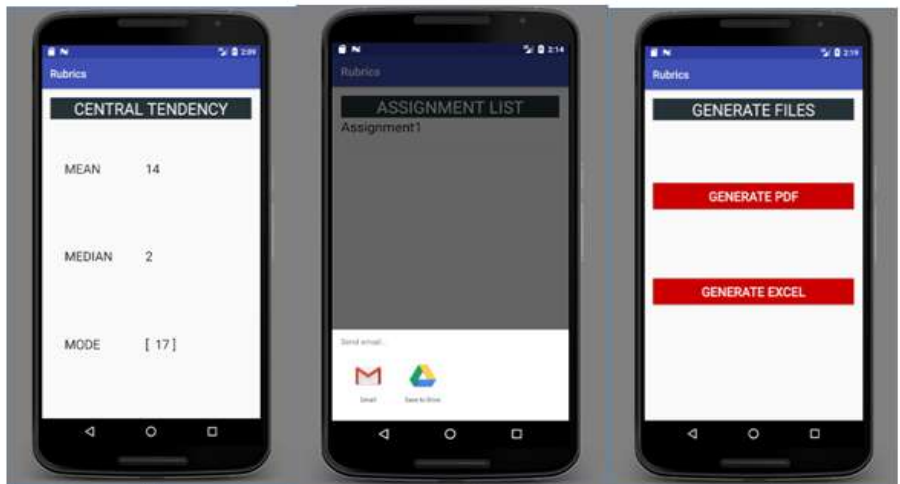

(j)       (k)       (l)

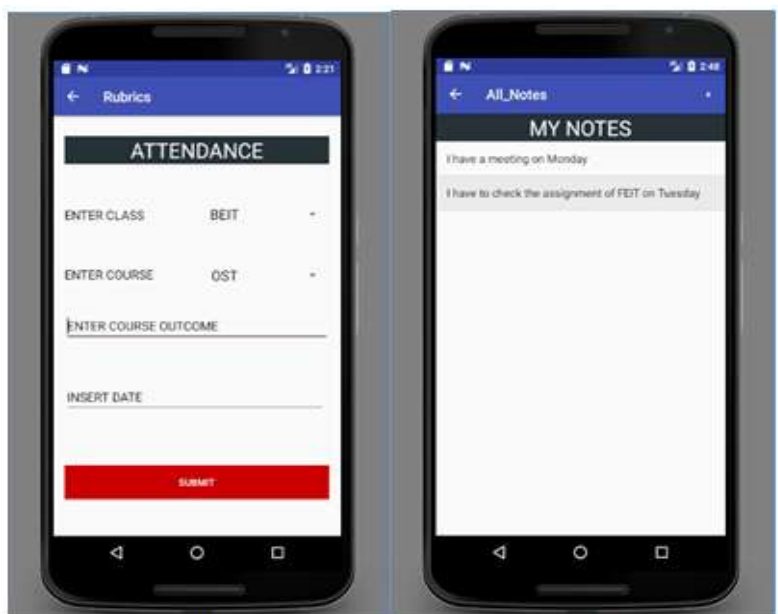

(m)       (n)

Fig. 8. User interfaces (a) Login (b) Register (c) Create class (d) Create class (e) Create course (f) Create assignment (g) Start grading (h) Total marks (i) Threshold marks (j) Central tendency (k) Student feedback (l) Generate files (m) Attendance (n) Notes

A run time analysis has been carried out based on the architecture discussed previously in Table 1. The table consists of output interfaces which are central tendency, graphs and total marks. As seen, the time taken for each task takes between 1 second to 1.5 seconds for the sample size from 50 to 150.

| SAMPLE SIZE | CENTRAL TENDENCY TIME(in seconds) | GRAPHS TIME(in seconds) | TOTAL MARKS TIME(in seconds) |
|---|---|---|---|
| 50 | 1 | 1 | 1 |
| 100 | 1 | 1.2 | 1.1 |
| 150 | 1.2 | 1.5 | 1.2 |

Table 1. Run time analysis with different sample sizes.

## V. CONCLUSION AND FUTURE OUTLOOK

This paper presents necessary guidelines about the Rubrics system for the mobile users who have installed the android application. Rubrics system is a mobile-based platform to the evaluator, wherein the evaluator will evaluate the students work according to the metric designed by the him/her. This system displays all the performance of the student in a graphical format so that evaluator will get a clear idea about the student's merits and lacunae as well the jobs that are pending and that which have been submitted. By using this system, the evaluator can review the student's work as well as give proper feedback to the them, who in turn get one-to-one attention from the evaluator. Future iteration of this application will work towards improving the feedback system in order to make it more personalised. Further as an advantage to the evaluator he/she would be able to judge the difficulty level of the question set by them based on the statistical tendencies given by the app. Thus, for the best learning as well as evaluation system Rubrics helps evaluators for fair evaluation and communicating with the students personally, thus providing proper guidance for the improvement of students.

**AUTHOR INFORMATION**

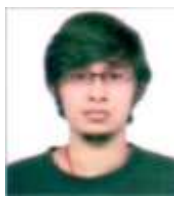
**Kaustubh Kundu** received his B.E degree in Information Technology from Don Bosco Institute of technology from University of Mumbai in 2017. He has done internship in IIT Bombay under Prof. Supratik Chakraborty. He is currently working as a Research assistant in IIT Bombay under Prof. Supratim Biswas.

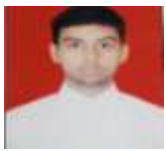
**Sushant Yadav** received his B.E degree in Information Technology from Don Bosco Institute of technology from University of Mumbai in 2017.

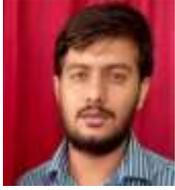 **Mr. Tayyabali Sayyad** is an Assistant Professor at Don bosco Institute of Technology. He has Master's in Computer Science and has 9 years of experience.